\documentclass{elsarticle}
\usepackage{amsmath}
\makeatletter
\@addtoreset{equation}{section}
\makeatother

\begin{document}
	\begin{frontmatter}
		
\title{Decoupling between gravitationally bounded systems and the cosmic expansion}
\author[1,2]{Yi Wang}
\ead{phyw@ust.hk}
\author[1,2]{Zun Wang}
\ead{zwangdq@connect.ust.hk}
	
\address[1]{Department of Physics, The Hong Kong University of Science and Technology, Clear Water Bay, Kowloon, Hong Kong, P.R.China}
\address[2]{Jockey Club Institute for Advanced Study, The Hong Kong University of Science and Technology, Clear Water Bay, Kowloon, Hong Kong, P.R.China}

\begin{abstract}

Recently, it was hypothesized that some supermassive black holes (SMBHs) may couple to the cosmic expansion. The mass of these SMBHs increase as the cubic power of the cosmic scale factor, leaving the energy density of the SMBHs unchanged when the universe expands. However, following general principles of general relativity, namely, locality or junction conditions, we show that the inner part of a gravitationally bounded system is unaware of the cosmic expansion, since the outer solution can be smoothly replaced by an asymptotically flat background. For the same reason, the direct reason that we do not expand with the expansion of the universe is not that we are bound states, but rather we are positioned in a greater gravitationally bounded system, namely the local group.	
\end{abstract}
\end{frontmatter}
\newpage


\section{Introduction}
Since the late 1990s, accelerated cosmic expansion has been indicated from supernova observations \cite{S. Perlmutter, A.G. Riess}, implying that the current universe is dominated by an energy component with constant density and negative pressure, dubbed dark energy (DE). In the last decades, numerous DE candidates have been come up with, such as (i) the cosmological constant by the quantum vacuum fluctuations, but the value of such vacuum energy predicted by most theories is 120 orders of magnitude larger than DE, known as a fine-tuning problem; (ii) Supersymmetry. However, supersymmetry must be broken below the scale of the particle theory Standard Model, still leaving a fine-tuning of more than 50 orders of magnitude; (iii) Scalar field theories,  such as slowly rolling scalar fields. However, for these mechanisms to work, one still has to explain why the bare cosmological constant is small. See, for example \cite{Li:2011sd} for a review of these proposals and references therein. 
 
In the recent literature, black holes (BHs) are considered as one new candidate as the source of DE \cite{Farrah1, Farrah2}. From observations, the mass of some BHs increases with the volume of the cosmic expansion \cite{Croker}. It was argued that this may not be a coincidence of black hole local accretion rate, but instead the expansion of BHs themselves coupling to the cosmic expansion, leading to a constant energy density just as DE. While this appears to produce the correct equation of state of DE, it also faces several doubts. For instance, the cosmological coupling hypothesis is based only on statistical fitting results from existing observations, with no theoretical support to explain the coupling mechanism. Furthermore, it is not clear on how BHs can contribute enough $\Omega_{\Lambda} \sim 0.7$ DE components, or how their perturbations can maintain stability under negative pressure. Also, the original cosmological constant problem of why the cosmological constant is small, is left unexplained.

In addition to these doubts, a more direct issue is that, in general relativity, the evolution of a system is only affected by the Riemann tensor where the system is located, but not affected by the spacetime curvature elsewhere. Within a gravitational bound state, say, a galaxy cluster or our local group, the local Riemann tensor is not the one derived from the FRW metric, but rather the one following from the gravitational bound state conditions. Thus, expansion should not manifest itself in galaxy clusters or our local group even for separation between free particles. To show this explicitly, here we show that gravity deep inside bound states can be described by a local effective theory, which do not exhibit the effects of cosmic expansion. This applies to the earth-moon system, the solar system, or the interior of galaxies. Therefore, BHs in galaxies, especially SMBHs at the cores of galaxies, should only be subject to local gravitational effects and cannot perceive the external expansion background, or even couple with it. This holds unless the locality of general relativity is fundamentally broken, invaliding junction conditions to connect solutions smoothly.

In this article, we investigate spherically symmetric gravitational solutions in the Friedmann-Robertson-Walker (FRW) background \cite{FRW1}. Using the McVittie solution \cite{McVittie1, McVittie2}, we can clearly discern the equilibrium condition between the effects of the gravitational force and cosmic expansion. Due to the extremely small cosmological constant $\Lambda$, the dominant gravitational effect is significantly enhanced by $r^3$ close to the system, completely canceling out the background expansion. This indicates that gravitationally bounded systems are not affected by the cosmic expansion background. To further verify this, we construct models to connect the spherically symmetric solutions with the FRW background using the timelike hypersurface shell $\Sigma$. Using junction conditions \cite{j1, j2}, we determine that the junction $\Sigma$ corresponds to the region where gravity and cosmic expansion are in equilibrium. In the local gravitational system, the Hubble parameter is not determined by the background, which violates the mechanism of cosmological coupling.

Our paper is organized as follows: In section 2, we provide a brief review of spherically symmetric solutions. In sections 3 and 4, we use a timelike shell hypersurface to separate the interior spacetime from the exterior cosmic background, thereby illustrating the lack of correlation between their respective Hubble parameters. Finally, in section 5, we discuss our findings and provide a summary of our results.

For the convention, we use $(+, -, -, -)$ and Planck units $(8 \pi G = c = \hbar = k_B = 1)$ through this paper for convenience.

\section{McVittie solution}
In general relativity, we care about the spherical symmetric solution to describe the spacetime outside the a star, galaxy cluster and so on. The detailed solutions are summarized in Appendix A. Here, we introduce the McVittie solution that represents such spherical gravitationally bounded system embedded in an FRW background. Based on the Schwarzschild solution, the metric reads
\begin{align}
ds^2 &= \left(\frac{1 - \frac{M}{2 a(t) r} \sqrt{1 + \kappa r^2}}{1 + \frac{M)}{2 a(t) r} \sqrt{1 + \kappa r^2}}\right)^2 dt^2 \\
& - \left(\frac{a(t)}{1 + \kappa r^2}\right)^2 \left(1 + \frac{M}{2 a(t) r} \sqrt{1 + \kappa r^2}\right)^4 (dr^2 + r^2 d\Omega^2)~,
\end{align}
where $M$ refers to the central object's mass, $a(t)$ is the scale factor and $d\Omega^2 = d\theta^2 + \sin^2\theta d\phi^2$. The parameter $\kappa \in \{-1, 0, 1\}$ is related to the 3-space curvature of the FRW spacetime.

For simplification, we consider the asymptotically spatially flat case with $\kappa = 0$ which fits most with the current cosmological data \cite{k1}. And we could define a time-dependent mass parameter
\begin{align}
\label{eq: mp}
\mu(t) = \frac{M}{a(t)}~.
\end{align}
Notice that the physical mass $M$ is fixed and won't be diluted by the universe expansion. The mass parameter $\mu(t)$ represents the mass inside the radius $r_c = a(t) r$ at time $t$ in the comoving frame. Then the McVittie solution becomes
\begin{align}
\label{eq: Mc2}
ds^2 = \left(\frac{1 - \mu(t)/2r}{1 + \mu(t)/2r}\right)^2 dt^2 - a^2(t) \left(1 +  \frac{\mu(t)}{2r}\right)^4 (dr^2 + r^2 d\Omega^2)~.
\end{align}
We could obtain the Schwarzschild solution and FRW metric directly from the McVittie solution \eqref{eq: Mc2} in the static spacetime $d a(t)/dt = 0$ and the massless case $\mu(t) = 0$. With a small mass parameter $\mu(t) \ll r$ by the universe evolution, it leads to a perturbed FRW solution with $ \phi = 2\mu(t)$ perturbation potential.

In Croker et al. \cite{Croker}, one of the main ideas to attribute the source of DE to BHs is a parameterized BH model with time-dependent mass in terms of the scale factor $a(t)$, that
\begin{align}
m(a) := m (a_0) \left(\frac{a}{a_0}\right)^k \ \ \ (a \geq a_0)~,
\end{align}
where $m(a)$ is the BH mass and $m(a_0)$ is the mass of the BH at its formation with the scale factor $a_0$. One question arises that what this mass $m(a)$ refers to. In the current mechanism, the physical mass can't be affected by the cosmic expansion with $k = 0$. If it refers to the mass from the comoving observers which is estimated by the astronomical observations, the parameter $k$ should be around $-1$ by \eqref{eq: mp}.

However, in Farrah et al. \cite{Farrah1, Farrah2}, their observations show that the offsets in the mass of SMBHs increase apparently compared with the ones in stellar mass reaching a factor of 7 between $z \sim 1$ and $z \sim 0$, and factor of 20 between $z \sim 2$ and $z \sim 0$. At $90 \%$ confidence, 
\begin{align}
k = 2.96_{-1.46}^{+1.65}\ \ \  \mathrm{and}\ \ \ k= 3.11_{-1.33}^{+1.19}~,
\end{align}
and excludes the $k = 0$ case at $99.98\%$ confidence. This indicates that the fitting result is not in agreement with any predictions for BH mass. It is essential to figure out the mechanism of the ``cosmological coupling'' and resolve its contradictions with solutions from the general relativity.

More intuitively, in general relativity, only the local Riemann tensor could affect the evolution of the system. This means that for one SMBH in the galaxy, it could only be affected by its own spacetime curvature, but cannot perceive the outside cosmic expansion, or even couple with it. This will be discussed more in the following section.

\section{Schwarzschild solution in FRW background}

The effect of the cosmic expansion is that two objects move away from each other radially over time, which could be equivalent to one type of repulsive force. It means that there will be some regions of the equilibrium between the gravity and such repulsive force by the cosmic expansion. We can intuitively figure it out from the asymptotically spatially flat McVittie solution \eqref{eq: Mc2} in the Schwarzschild coordinates that 
\begin{align}
\label{eq:egce}
x \nonumber &= r \left(1 + \frac{M}{2 a(t) r}\right)^2  a(t)\ \ \Rightarrow \\
ds^2 &= \left(1 - \frac{2M}{x} - H^2(t) x^2\right)dt^2 + \frac{2 H(t) x}{\sqrt{1 - 2M / x}} dx dt - (1 - \frac{2M}{x})dx^2 - x^2 d\Omega^2~,
\end{align}
where $H(t) = \dot{a} /a $ is the Hubble parameter. Considering a small $x$ in the local frame, the gravity potential term $2M/ x$ will be the dominated effect in the rate of $x^3$ compared with such small value of today's Hubble parameter $H \sim 2.19 \times 10^{18} \mathrm{(s^{-1})}$. For the equilibrium region between the gravity and cosmic expansion with radius $R_e$, we have 
\begin{align}
\label{eq: eqp}
\frac{2M}{R_e} = H^2 (t) R_e^2 \ \ \Rightarrow \ \ R_e = \left(\frac{2M}{H^2(t)}\right)^{1/3}~.
\end{align}
Within this region with $R_e$, the local gravitationally bounded system cannot be affected by the outside cosmic background.
\\

\begin{figure}[h]
    \centering
    \includegraphics[width=0.7\textwidth]{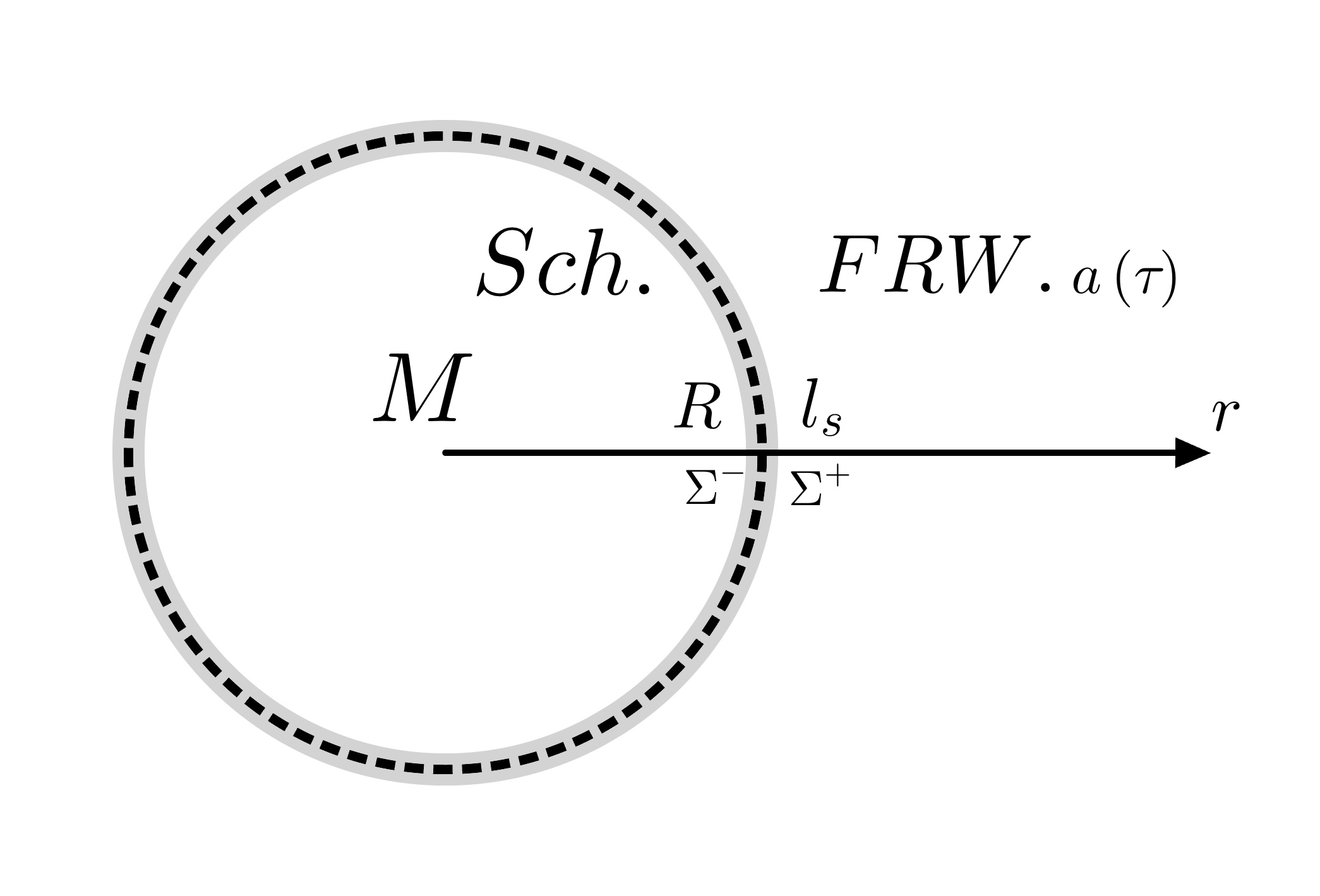}
    \caption{The joint metric with inside Schwarzschild solution and outside FRW background by junction $\Sigma$. The massless junction shell comoves with the background, but the inside can't perceive the outside cosmic expansion.}
    \label{fig:Schvs}
\end{figure}

Next, we construct one joint metric to show this picture more concretely as shown in Figure.\ref{fig:Schvs}. One timelike hypersurface $\Sigma$ is used to glue two different solutions to form the new joint solution to describe the whole spacetime. The inside solution $g^-$ is the Schwarzschild solution with central fixed mass $M$ and the outside solution $g^+$ refers to the FRW background. Then the junction $\Sigma = \Sigma^- \cap \Sigma^+$ should be a dynamical shell between them. 
The inside junction $\Sigma^-$ in the comoving coordinates $(t, r, \theta, \delta)$ is
\begin{align}
ds_-^2 = \left(1 - \frac{2 M}{R}\right)  d t^2 - \frac{d r^2}{\left(1 - \frac{2 M}{R}\right)} - R^2 d\Omega^2~,
\end{align}
and the outside junction $\Sigma^+$ in FRW spacetime with coordinates $(\tau, l, \theta, \delta)$ behaves as
\begin{align}
ds_+^2 = d\tau^2  - a^2(\tau) l_s^2 d\Omega^2
\end{align}
in FRW spacetime with the scale factor $a(\tau)$. The junction shell $\Sigma$ is massless which means that it should be in sync with the FRW background with fixed value $l = l_s$ at $\Sigma$, and for the inside solution, it match with the boundary at $r = R$. We summarize more details on junction conditions calculation in Appendix B. Follow the junction conditions, the joint metric exists with such conditions
\begin{align}
R & = \left(\frac{2 M}{H^2 (\tau)}\right)^{1/3}\\
t &= \int_{\tau_0}^{\tau} (1 - H^2 R^2)^{-1} d\tau~.
\end{align}
We notice that the position of the junction $R(\tau)$ just match with the equilibrium region between gravity and cosmic expansion in \eqref{eq: eqp}
\begin{align}
\label{eq:rr}
R(\tau)= R_e (\tau) = \left(\frac{2 M}{ H^2 (\tau)}\right)^{1/3}~,
\end{align}
which consistent with the conclusion by the locality of general relativity. The position of the junction shell $\Sigma$ depends on the Hubble parameter of FRW background, but within the junction, the inside spacetime curvature is only affected by the gratatinally bounded system itself. If we place a pair of test particles in the interior spacetime, they will only perceive the local Schwarzschild background and no cosmic expansion effect will be detected. For the case where the internal spacetime is also an expansion solution, we will discuss this in the next section.

\section{McVittie solution in FRW background}

In this section, we replace the interior spacetime with McVittie solution which Hubble parameter could be different from the one of FRW background. A timelike junction shell junction $\Sigma$ intersects with both solution at $\Sigma^{\pm}$. The inside McVittie solution is described with scale factor $b(t)$ in comoving coordinates
$(t, r, \theta, \delta)$. For the outside FRW background, the scale factor is still $a(\tau)$ with $(\tau, l, \theta, \delta)$ coordinates.
\begin{figure}[hh]
    \centering
    \includegraphics[width=0.7\textwidth]{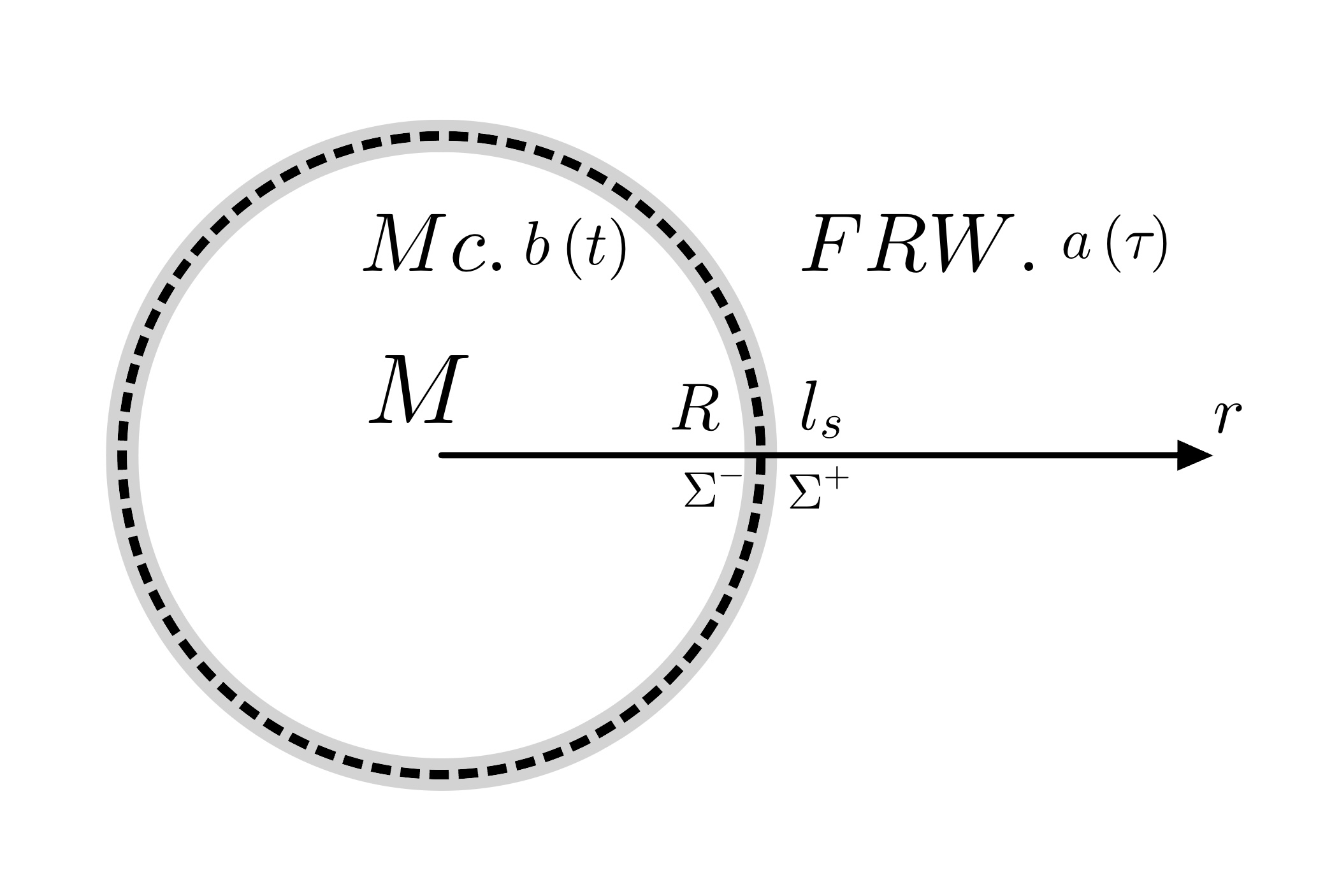}
    \caption{The joint metric with inside McVittie solution and outside FRW background. The inside scale factor is $b(t)$, which could be different from the outside one $a(\tau)$. The inside gravitationally bounded system still couldn't perceive the outside cosmic expansion even though the inside spacetime is also the expanding solution.}
    \label{fig:Mcvs}
\end{figure}
The McVittie solution could be written in the isotropic form as
\begin{align}
ds_-^2 = \left(\frac{1 - \frac{M}{2 R b(t)}}{1 + \frac{M}{2 R b(t)}}\right)^2 dt^2 - b^2(t)\left(1 + \frac{M}{2 R b(t)}\right)^4 (dr^2 + R^2 d\Omega^2)~.
\end{align}
On the outside
\begin{align}
ds_+^2 = d\tau^2 - a^2(\tau) l_s^2 d\Omega^2~.
\end{align}
Follow the junction conditions at $\Sigma$, the matched metric reads
\begin{align}
 & R^2 b^2(t) \left(1 + \frac{M}{2 R b(t)}\right)^2 = a^2(\tau) l_s^2\\
& \left(\frac{1 - \frac{M}{2 R b(t)}}{1 + \frac{M}{2 R b(t)}}\right)^2 \dot{t}^2 - \dot{R}^2 b^2(t) \left(1 + \frac{M}{2 R b(t)}\right)^4 - 1 = 0~,
\end{align}
and $K_{\mu \nu}^- = K_{\mu \nu}^+$ must be satisfied, where any functions satisfy $\dot{A} \equiv dA/ d\tau$ for convenience. Then we could figure out the position of such junction should satisfy
\begin{align}
R = \left(\frac{2 M }{b(t) H^2 (\tau)}\right)^{1/3}~,
\end{align}
which depends on the interior physical mass and the scale factors for two sides. This also matches with the equilibrium region between potential terms of gravity and cosmic expansion in McVittie solution just like \eqref{eq:rr}. In particular, we could let $b(t) = 1$ to simplify this to the case in the last section. This means that the joint metric satisfying Einstein field equation could have a nonuniform distribution of the Hubble parameters. Even though the interior spacetime is also an expanding solution, the gravitationally bounded system still cannot perceive the outside cosmic expansion.

\section{Discussion and summary}
We considered two interior spherically symmetric solutions to be joined with the cosmic background, following the corresponding junction conditions for the joint solution. The junction $\Sigma$ is dynamically determined by the Hubble parameters of both sides and indicates the equilibrium region between gravity and cosmic expansion. For the interior static Schwarzschild solution combined with the FRW background, such an equilibrium region is related to
\begin{align}
R_e \sim \left(\frac{2 M}{H^2(\tau)}\right)^{1/3}~.
\end{align}
Our results show that local gravitationally bounded systems cannot be affected by the cosmic background. For BHs within galaxies, they cannot perceive the outside cosmological expansion. On scales larger than that between galaxies, gravity is weak enough for the cosmic expansion to manifest.

Our calculation, though simple, clarifies some issues that was usually not properly interpreted. For example, why our body, or the solar system does not expand with the cosmic expansion? The conventional explanation is that they are bounded systems and the cosmic expansion is too weak to separate them. However, our work shows that this is not the direct reason\footnote{What is a ``direct reason''? An analogy may be helpful here: imagine that we are sitting in a house. Outside the house (and only outside the house), wind is blowing strongly. Why aren't we blown away by the wind? The direct reason should be that, there is no wind in the house at all, instead of calculating the fraction force to show that it can balance the force by the wind if we were outside the house. }. Since us, and the solar system are deep inside our local group, a gravitationally bounded system, even if our body or the solar system were not bounded by additional forces but rather free particles, they will not expand with the expansion of the universe since these free particles do not perceive cosmic expansion at all. In explaining the decoupling from cosmic expansion, only the largest gravitationally bounded system matters.

Having that said, if further observations confirm that the mass of BHs indeed increases in coupling to the cosmic expansion, instead of coincidentally a similar accretion rate, we need to understand how this coupling fits into the theory of gravity, i.e. how general relativity should be modified for this to happen.

\section*{Acknowledgement}
We thank Tingqi Cai, Chao Chen, Jinbo Jiang, Tao Liu and Mian Zhu for helpful discussions and comments. This work was supported in part by the National Key R$\&$D Program of China (2021YFC2203100), the NSFC Excellent Young Scientist Scheme (Hong Kong and Macau) Grant No.~12022516, and by the RGC of Hong Kong SAR, China Grant No.~16306422.

 \appendix
\section{Detailed solutions of spherically symmetric distribution}
\subsection{Schwarzschild solution}
The static spherically symmetric metric is
\begin{align}
\label{eq:sssm}
ds^2 = e^{\nu(r)} dt^2 - e^{\lambda(r)} dr^2 - r^2(d \theta^2 + \sin^2{\theta} d\phi^2)~,   
\end{align}
where the coefficients $\nu(r)$ and $\lambda(r)$ only depends on $r$.

Then, we consider a spherically symmetric galaxy cluster composed by non-relativistic gas without charge and spin, where the energy density $\rho = \rho(r)$ and pressure $p = p(r)$ are only functions of $r$ with equation of state $p = p(\rho)$.

The Einstein field equations in the static spacetime
\begin{align}
R_{\mu \nu} - \frac{1}{2} R g_{\mu \nu} = T_{\mu \nu}~,
\end{align}
where $T_{\mu \nu}$ refers to the energy-momentum tensor assumed to be the form of the perfect fluid tensor with 4-velocity $u_{\mu}$ as $T_{\mu \nu} = \rho u_{\mu} u_{\nu} + p(u_{\mu} u_{\nu} - g_{\mu \nu})$. Only consider the nonzero terms on the diagonal, we could get three independent nonlinear differential equations,
\begin{align}
\rho + 3p &= e^{-\lambda} \left[\nu'' + \frac{1}{2}\nu' \left(\nu' - \lambda' + \frac{4}{r}\right)\right]\\ 
\rho - p &= e^{-\lambda} \left[-\nu'' - \frac{1}{2} \nu' \left(\nu' - \lambda'\right) + \frac{2 \lambda'}{r}\right]\\
\rho - p &= e^{-\lambda}\left[ \frac{1}{2 r}(\lambda' - \nu') - \frac{1}{r^2}\right] + \frac{1}{r^2}~,
\end{align}
where any functions satisfies $A' \equiv \partial A(r)/\partial r$ for convenience. Here, the fourth component equation on $T_{33}$  is ignored due to the similarity to $T_{22}$ component with additional $\sin^2{\theta}$ coefficient. Then we could get the expressions of energy density $\rho(r)$ and the pressure $p(r)$ respectively
\begin{align}
\label{eq:rhop}
\rho(r) &= \frac{e^{\lambda}}{r^2}\left(\lambda' r - 1\right) + \frac{1}{r^2}\\
p(r) &= \frac{e^{-\lambda}}{r^2}\left(\nu' r + 1\right) - \frac{1}{r^2}~,
\end{align}
and one constraint equation on $\lambda$ and $\nu$ 
\begin{align}
\frac{e^{\lambda}}{r^2} = \frac{1}{2}\left[ -\nu'' + \frac{(\lambda' + \nu')}{r} + \frac{1}{2}\nu'(\lambda' - \nu') + \frac{2}{r^2}\right]~.  
\end{align}
Consider a bounded and constant energy density distribution for simplification, this should describe the exterior Schwarzschild metric. So we could follow that form to define a general mass function $m = m(r)$, which satisfies $g_{rr} \equiv \left(1 - \frac{2 m(r)}{r}\right)^{-1}$. Then from \eqref{eq:rhop} we could get the expression of the solution of spherically symmetric distribution with given $\rho(r)$ and $p(r)$,
\begin{align}
\label{eq:ggmetric}
g_{rr} &= -e^{\lambda} = - \left(1- \frac{2 m(r)}{r}\right)^{-1}\\
g_{tt} &= e^{\nu} = \exp{\left[\int_0^r \left(\frac{2 m(r') + p(r') r'^3}{r'^2 - 2 m(r') r'}\right) dr'\right]}~,
\end{align}
where the mass function is defined as
\begin{align}
m(r) = \int_0^r \frac{1}{2} \rho(r') r'^2 dr'~.
\end{align}
For the special case, if such galaxy cluster is only bounded in $0<r<R$, this could lead to the Schwarzschild metric in Schwarzschild coordinates $(t, r, \theta, \phi)$ with fixed mass function $m(r) = M$. For the interior solution $(0<r<R)$, its line element is
\begin{align}
\label{eq:isch}
ds^2 \nonumber &= \frac{1}{4}\left(3 \sqrt{1- \frac{2M}{R}}
- \sqrt{1 - \frac{2 M r^2 }{R^3}}\right)^2 dt^2 \\
&- \left(1-\frac{2M r^2}{R^3}\right)^{-1}dr^2 - r^2(d\theta^2 + \sin^2\theta d\phi^2)~.
\end{align}
For the exterior solution $(r>R)$, the line elements becomes
\begin{align}
\label{eq:esch}
ds^2 = \left(1 - \frac{2M}{r}\right)dt^2 - \left(1 - \frac{2M}{r}\right)^{-1} dr^2 - r^2(d\theta^2 + \sin^2\theta d\phi^2)~,
\end{align}
which matches with the interior solution at the surface $r =R$ with the same line element \cite{sch1, sch2}. Furthermore, the junction conditions need to be applied to match their extrinsic curvature.
\subsection{Solutions in isotropic coordinates}
Here we derive the form of isotropic coordinate for solutions under spherically symmetric distribution. Like \eqref{eq:sssm}, the metric could be defined as
\begin{align}
ds^2 = e^{\alpha(r)} dt^2 - e^{\beta(r)}(dr^2 + r^2 d\theta^2 + r^2 \sin^2\theta d \phi^2)~,
\end{align}
where parameters $\alpha(r)$ and $\beta(r)$ are only functions of $r$.
From the Einstein field equation, we could get three independent nonlinear differential equations
\begin{align}
\rho + 3p &= e^{-\beta} \left(\alpha'' + \frac{\alpha'^2}{2} + \frac{\alpha' \beta'}{2} + \frac{2 \beta'}{r}\right)\\
\rho - p &= e^{-\beta} \left(-2\beta'' - \alpha'' - \frac{\alpha'^2}{2} + \frac{\alpha' \beta'}{2} - \frac{2\beta'}{r}\right)\\
\rho - p &= -\frac{1}{2}e^{-\beta}\left(2 \beta'' + \frac{2\alpha'}{r} + \beta'^2 
+ \alpha'\beta' + \frac{6 \beta'}{r}
\right)~.
\end{align}
Then the energy density $\rho(r)$ and pressure $p(r)$ could be expressed as
\begin{align}
\rho(r) &= -e^{-\beta} \left( \beta'' + \frac{2\beta'}{r} + \frac{\beta'^2}{4}\right)\\
p(r) &= e^{-\beta} \left(\frac{\alpha'}{r} + \frac{\beta'}{r} + \frac{\alpha' \beta'}{2} + \frac{\beta'^2}{4}\right)~.
\end{align}
Notice that the energy density is on parameter $\beta(r)$ from the differential equation. If we consider the Schwarzschild case with the constant and bounded $(0<r<R)$ energy density distribution without the singularity, the general solution is
\begin{align}
\beta = \log{\left[\frac{12 }{\rho (A r^2 + A^{-1})^2}\right]}~.
\end{align}
And from the constraint by A.3 and A.4, the general form of $\alpha$ should be 
\begin{align}
\alpha = 2\log{\left[\frac{1}{(1+A^2 r^2)} - B\right]^2} + C~,
\end{align}
where $A$, $B$ and $C$ could be arbitrary constant for the above expressions which could be evaluated by the given boundary conditions. At the distribution boundary $r = R$, the pressure should be zero $p(R) = 0$ and $g_{\mu \nu}$, $\partial g_{\mu \nu}$ should be continuous here. Finally, we could get the exterior isotropic solution $(r>R)$
\begin{align}
ds^2 = \left(\frac{1 - m/2r}{1 + m/2r}\right)^2 dt^2 - (1 + m/2r)^4 dr^2 - r^2 d\Omega^2~,
\end{align}
and the interior isotropic solution $(0<r<R)$
\begin{align}
ds^2 \nonumber &= \left[\frac{m^2 r^2 + 4 R^4 - 4 m R(r^2 + R^2)}{(m+2R)(mr^2 + 2R^3)}\right]^2 dt^2\\
& - \left[\frac{(m + 2R)^3}{8(mr^2 + R^3)}\right]^2 dr^2 -r^2 d\Omega^2~,
\end{align}
which match with each other at the boundary $r= R$.
\section{Junction conditions}
In the previous literature, there are several junction conditions with different degrees of constraints. In this work we choose the commonly used Israel junction conditions to glue two spacetime solutions \cite{BHU}. The requirement is that their metric and the extrinsic curvature $K$ should match with each other at the junction $\Sigma$, where $K$ is defined as
\begin{align}
K_{\mu \nu} = (n_c \Gamma^c_{a b} - \partial_a n_b)e^a_{\mu} e^b_{\nu}~,
\end{align}
where $e^{a}_{\mu} = \partial x^a/ \partial y^{\mu}$ and $n$ refers to the normal vector to the junction.

There are connections between the inside spacetime and the FRW comoving time $\tau$ with $t = t(\tau)$ and $r = r(\tau)$ so the inside one $\Sigma^-$ could be in the same form as $\Sigma^+$ that
\begin{align}
ds_-^2 = \left[\left(1 -\frac{2 M}{r}\right) \dot{t}^2 - \frac{\dot{r}^2}{\left(1 -\frac{2 M}{r}\right)}\right]d\tau^2 - r^2 d\Omega^2~.
\end{align}
At the junction shell, the first condition requires
\begin{align}
\label{eq:b3}
R(\tau) &= a(\tau) l_s\\
\left(1-\frac{2M}{R}\right)^2 \dot{t}^2 & - \left(1-\frac{2M}{R}\right) - \dot{R}^2 = 0~.
\end{align}
For the second condition to match the extrinsic curvature normal to $\Sigma$, we only care about the independent $K_{\tau \tau}$ and $K_{\theta \theta}$ components since $K_{\delta \delta} = \sin^2\theta K_{\theta \theta}$. For the inside junction $K^-$ in Schwarzschild solution,
\begin{align}
K^-_{\tau \tau} \nonumber &= \sum (n_c \Gamma_{a b}^c - \partial_a n_b)e_{\tau}^a e_{\tau}^b\\
&= \frac{2 M \dot{t}}{R^2} + \frac{\ddot{t}}{\dot{R}}\left(1 - \frac{2 M}{R}\right) + \frac{(\ddot{R} + M/ R)}{\dot{t} \left(1 - \frac{2 M}{R}\right)}~,
\end{align}
\begin{align}
\label{eq:b6}
K^-_{\theta \theta} &= \partial_{\tau} t \Gamma_{\theta \theta}^r = - \left(1 - \frac{2 M}{R}\right) R \dot{t}~.
\end{align}
And for the outside one $K^+$ in FRW spacetime,
\begin{align}
K^+_{\tau \tau} &= (a \Gamma_{\tau \tau}^l \partial_{\tau} n^+_{\tau}) e_{\tau}^{\tau} e_{\tau}^{\tau} = 0\\
K^+_{\theta \theta} &= (a \Gamma_{\theta \theta}^l - 0)e_{\theta}^{\theta} e_{\theta}^{\theta}= - a l_s = -R~.
\end{align}
Under the requirement $K_{\mu \nu}^- = K_{\mu \nu}$, from \eqref{eq:b3} to \eqref{eq:b6} with four independent equations, we could get
\begin{align}
R(\tau)&= \left(\frac{2 M}{H^2}\right)^{1/3}\\
t(\tau)&=\int_{\tau_0}^{\tau} (1 - H^2 r^2)^{-1} d\tau~,
\end{align}
where $H = \dot{a}/a(\tau)$ refers to the Hubble parameter outside.

\end{document}